\pdfoutput=1
\def\mode{1}
\documentclass[sigconf, natbib=true, screen=true, 10pt, authorversion=true]{acmart} %

\setcopyright{acmlicensed}

\copyrightyear{2020}
\acmYear{2020}
\acmConference[SIGMOD'20]{Proceedings of the 2020 ACM SIGMOD International Conference on Management of Data}{June 14--19, 2020}{Portland, OR, USA}
\acmBooktitle{Proceedings of the 2020 ACM SIGMOD International Conference on Management of Data (SIGMOD'20), June 14--19, 2020, Portland, OR, USA}
\acmPrice{15.00}
\acmDOI{10.1145/3318464.3383132}
\acmISBN{978-1-4503-6735-6/20/06}

\title{Optimal Join Algorithms Meet Top-\k}

\author{Nikolaos Tziavelis}
\affiliation{%
    \orcidicon{0000-0001-8342-2177}
    \institution{Northeastern University}
    \city{Boston}
    \state{Massachusetts}
    \country{USA}
}
\email{ntziavelis@ccs.neu.edu}

\author{Wolfgang Gatterbauer}
\orcid{0000-0002-9614-0504}%
\affiliation{%
    \orcidicon{0000-0002-9614-0504}
    \institution{Northeastern University}	
    \city{Boston}
    \state{Massachusetts}
    \country{USA}
}
\email{w.gatterbauer@northeastern.edu}

\author{Mirek Riedewald}
\affiliation{%
    \orcidicon{0000-0002-6102-7472}
    \institution{Northeastern University}
    \city{Boston}
    \state{Massachusetts}
    \country{USA}
}
\email{m.riedewald@northeastern.edu}

\usepackage{url}
\usepackage{enumitem} %

\usepackage{upgreek} %

\usepackage{listings} %

\usepackage{tabularx} %
\usepackage{booktabs} %
\usepackage{multirow}
\usepackage{hhline} %
\usepackage{pgfplotstable} %

\usepackage{easybmat} %

\usepackage{subcaption} %
\usepackage{rotating}		%

\usepackage{xspace} %

\usepackage[ruled,noend,linesnumbered]{algorithm2e} %

\DontPrintSemicolon     %
\SetNlSty{}{}{}                %
\SetAlgoInsideSkip{smallskip}   %
\SetAlCapSkip{1.5mm}             %
\SetInd{1.5mm}{1.5mm}             %

\SetAlFnt{\small}			%
\SetAlCapFnt{\small}		%
\SetAlCapNameFnt{\small}

\newtheorem{questionW}{Question}
\newtheorem{resultW}{Result}

{\end{itshape}
}
\let\footnotesize\scriptsize

\newcommand{\hide}[1]{}

\usepackage{hyperref}   %
\hypersetup{                                                            %
	pdfpagemode=UseNone,               %
    pdfstartview=Fit,
    pdfpagelayout=SinglePage,       %
    bookmarksnumbered=true,
    bookmarksopen=false,             %
    unicode,
    colorlinks=true,       %
    linkcolor=blue,          %
    citecolor=cyan,  %
    filecolor=magenta,      %
}

\usepackage[capitalise,nameinlink]{cleveref} %

\crefname{algocf}{alg.}{algs.}
\Crefname{algocf}{Algorithm}{Algorithms}
\crefalias{AlgoLine}{line}%
\if 1\mode
	\AtEndPreamble{%
	    \hypersetup{colorlinks,
	      linkcolor=purple,
	      citecolor=blue,
	      urlcolor=ACMDarkBlue,
	      filecolor=ACMDarkBlue}}
\fi

\if 0\mode
	\makeatletter 

	\renewcommand{\@opargbegintheorem}[3]{%
	    \parskip 0pt %
	    \trivlist
	    \item[%
	    	\hskip 10\p@										%
	        \hskip \labelsep
	        {\sc #1\ #2\             %
	   \setbox\@tempboxa\hbox{(#3)}  %
	        \ifdim \wd\@tempboxa>\z@ %
	            \hskip 0\p@\relax    %
	            \box\@tempboxa       %
	        \fi.}%
	    ]
	    \it
	}

	\def\@begintheorem#1#2{%
	    \parskip 0pt %
	    \trivlist
	    \item[%
	    	\hskip 10\p@										%
	        \hskip \labelsep
	        {{\sc #1}\hskip 5\p@\relax#2.}%
	    ]
	    \it
	}

	\makeatother
\fi

\usepackage{tcolorbox}		%
\tcbuselibrary{breakable,skins}		%
\tcbset{
		enhanced jigsaw,	%
		colback=gray!20,
		colframe=gray!40,
		arc=0mm,
		boxrule=1pt,		%
		left=1pt,
		right=1pt,
		topsep at break=1mm,			%
		top=1pt,		%
		bottom=0mm,		%
		breakable,		%
		parbox = false		%
		}

\makeatletter
\DeclareRobustCommand*\uell{\mathpalette\@uell\relax}
\newcommand*\@uell[2]{
  \setbox0=\hbox{$#1\ell$}
  \setbox1=\hbox{\rotatebox{10}{$#1\ell$}}
  \dimen0=\wd0 \advance\dimen0 by -\wd1 \divide\dimen0 by 2
  \mathord{\lower 0.1ex \hbox{\kern\dimen0\unhbox1\kern\dimen0}}
}
\makeatother

\newcommand{\smallsection}[1]{\vspace{2mm}\noindent\textbf{#1.}} %

\renewcommand{\epsilon}{\varepsilon} %
\renewcommand{\O}{{\mathcal{O}}} %

\newcommand{\NPRR}{\textsc{NPRR}\xspace}

\newcommand{\GenJoin}{\textsc{Generic-Join}\xspace}

\newcommand{\inp}{\mathrm{IN}}
\newcommand{\out}{\mathrm{OUT}}
\renewcommand{\inp}{\texttt{in}}
\renewcommand{\out}{\texttt{out}}
\renewcommand{\inp}{n}
\renewcommand{\out}{r}

\usepackage{booktabs} %
\graphicspath{{./figs/}}
\usepackage{numprint}
\usepackage{multirow}

\renewcommand{\k}{\textit{k}\xspace}

\usepackage{scalerel}
\usepackage{tikz}
\usetikzlibrary{svg.path}

\definecolor{orcidlogocol}{HTML}{A6CE39}
\tikzset{
  orcidlogo/.pic={
    \fill[orcidlogocol] svg{M256,128c0,70.7-57.3,128-128,128C57.3,256,0,198.7,0,128C0,57.3,57.3,0,128,0C198.7,0,256,57.3,256,128z};
    \fill[white] svg{M86.3,186.2H70.9V79.1h15.4v48.4V186.2z}
                 svg{M108.9,79.1h41.6c39.6,0,57,28.3,57,53.6c0,27.5-21.5,53.6-56.8,53.6h-41.8V79.1z M124.3,172.4h24.5c34.9,0,42.9-26.5,42.9-39.7c0-21.5-13.7-39.7-43.7-39.7h-23.7V172.4z}
                 svg{M88.7,56.8c0,5.5-4.5,10.1-10.1,10.1c-5.6,0-10.1-4.6-10.1-10.1c0-5.6,4.5-10.1,10.1-10.1C84.2,46.7,88.7,51.3,88.7,56.8z};
  }
}

\RequirePackage{etoolbox}
\DeclareRobustCommand\orcidicon[1]{\href{https://orcid.org/#1}{\mbox{\scalerel*{
\begin{tikzpicture}[yscale=-1,transform shape]
\pic{orcidlogo};
\end{tikzpicture}
}{|}}}}

\begin{document}

\begin{abstract}
\emph{Top-\k queries} have been studied intensively in the database community and they are an important
means to reduce query cost when only the ``best'' or ``most interesting''
results are needed instead of the full output. 
While some optimality results exist, e.g., the famous Threshold Algorithm,
they hold only in a fairly limited model of computation that does not account for the cost incurred by
large intermediate results and hence is not aligned with typical database-optimizer cost models.
On the other hand, the idea of avoiding large intermediate results is arguably the main goal of
recent work on \emph{optimal join algorithms}, which uses the standard RAM model of computation
to determine algorithm complexity. This research has created a lot of excitement due to its promise
of reducing
the time complexity of join queries with cycles, but it has mostly focused on full-output computation.
We argue that the two areas can and should be studied from a unified point of
view in order to achieve optimality in the common model of computation for a very general
class of top-\k-style join queries.
This tutorial has two main objectives.
First, we will explore and contrast the main assumptions, concepts, and algorithmic achievements of
the two research areas.
Second, we will cover recent, as well as some older, approaches that emerged at the intersection
to support efficient \emph{ranked enumeration of join-query results}.
These are related to classic work on \k-shortest path algorithms and more general optimization
problems, some of which dates back to the 1950s. We demonstrate that this line of research
warrants renewed attention in the challenging context of ranked enumeration for general
join queries.
\end{abstract}

\maketitle

\section{Introduction}

Join-query evaluation is a fundamental problem in databases, hence it is not surprising that recent
work on \emph{worst-case-optimal} (WCO) join algorithms~\cite{ngo2018worst,Ngo:2012:WOJ:2213556.2213565}
generated a lot of excitement. The basic insight is that standard join algorithms that treat multiway
joins with cycles as a sequence of pairwise joins are provably suboptimal in that they may
produce intermediate results that are asymptotically larger than the largest output
this query may produce over \emph{any possible input instance}.
By taking a ``holistic'' approach, WCO join algorithms guarantee a running-time complexity
that matches the worst-case output size of a given query~\cite{ngo2018worst}.
Interestingly, recent work on factorized databases~\cite{olteanu15dtrees} and
``optimal'' join algorithms~\cite{khamis17panda}\footnote{We will elaborate more on the distinction
between optimal and worst-case-optimal join algorithms in \cref{sec:optimal-join-algorithms}.}
has shown that the same and even better
time complexity can be achieved by decomposing a cyclic join query into multiple acyclic join plans and
routing different subsets of the input to different plans.
A key insight is that WCO join algorithms are \emph{not output sensitive}: their complexity
guarantees do not improve when a query has only a small output, e.g., when none of the input
tuples in a given database instance happen to form a result.
Similarly, the time-complexity guarantees of WCO join algorithms are weak in the presence
of projections (e.g., for Boolean join queries, which ask if the join has any result).

Since worst-case optimality 
is defined with respect to
the largest output
of the query over all possible inputs, it is not a natural
fit for top-\k queries, which aim to reduce query cost when only few results are needed.
Consider a graph with weighted edges, where lower weights represent greater importance, and
the problem of finding the top-\k \emph{lightest 4-cycles}, i.e.,
the \k most important cycles consisting of 4 edges.
This, as well as any other graph-pattern query, can be expressed with self-joins of the
edge set: here a 4-way join with equality conditions on the endpoints of adjacent edges.\footnote{For simplicity we ignore the issue of degenerate cycles, i.e., the same node or edge can appear
more than once in the cycle.}
Abstractly, all results are sorted according to a
\emph{ranking function} and the query needs to return only the first \k of them.
In a graph with $n$ edges, there can be $\O(n^2)$ 4-cycles, therefore a WCO join algorithm
would run in time $\O(n^2)$. On the other hand, 
it has been shown that
the corresponding Boolean
query (``Is there any 4-cycle?'') can be answered in $\O(n^{1.5})$~\cite{khamis17panda}.
It is tempting to assume that for small $k$, finding the \k lightest cycles will have complexity
close to the Boolean query, and as we will demonstrate this turns out to be
correct~\cite{Tziavelis:fullversion}.

Interestingly, the above question had not been addressed by the extensive literature on top-\k queries
in the database context~\cite{ilyas08survey,rahul19topk}. There exist approaches with optimality
guarantees, e.g., the Threshold Algorithm~\cite{fagin03}, but their optimality holds only
in a restricted model of computation where cost is measured in terms of the number of tuples accessed,
while the actual computation is essentially ``free.''\footnote{The original motivation for this model are
middleware settings where the algorithm is charged for requests made to external input sources.}
We instead will discuss and analyze all top-\k algorithms from the point of view of the standard
\emph{RAM model} of computation that charges $\O(1)$ for each memory access, i.e., it also accounts
for cost incurred by large intermediate results and agrees with the model used in the context of
(worst-case) optimal join algorithms.

This tutorial will generally survey these two seemingly different
areas---optimal joins and top-\k---from a unified point of view.
We intend to achieve this by highlighting the underlying assumptions made by illustrating
important achievements and algorithmic ideas in the two lines of work.
By  formally defining the common foundations, we are able to reveal fruitful research
directions at the intersection: How can we extend optimal join algorithms with ideas
from top-\k query processing to create frameworks for \emph{optimal ranked enumeration} over general
join queries? What types of ranking functions can be supported efficiently? And how can sorting
be pushed deep into the join computation?
While some recent work has started to explore those
questions~\cite{chang15enumeration,deep19,KimelfeldS2006,Tziavelis:fullversion,yang2018any,YangRLG18:anyKexploreDB},
much is still left to be done.

\smallsection{Audience}
The tutorial targets researchers and practitioners who desire an intuitive introduction
to recent developments in the theory of optimal join algorithms, including topics such as
generalized and fractional hypertree decompositions of cyclic queries, 
different notions of query width,
fractional edge cover,
factorized representation,
increasingly tight notions of optimality,
and enumeration algorithms for join queries.
It is also suitable for those interested in a concise comparison of major top-\k approaches
that were proposed in the context of join queries.

\smallsection{Prerequisites}
To make all material accessible to those interested in the practical impact of the techniques,
the tutorial will heavily favor intuitive examples and explanations over low-level technical
details. In the same spirit, and in line with much of the recent work on optimal join algorithms,
we generally take a database-centric view and will present asymptotic complexity results in terms of
\emph{data complexity} in $\tilde\O$-notation (read as ``\emph{soft}-$\O$'').
Data complexity treats query size (i.e., the size of the query expression itself)
as a constant and focuses on scalability in the size of the data.
The $\tilde\O$-notation abstracts away poly-logarithmic factors in input size as those factors often
clutter a formula and poly-log grows asymptotically slower than a linear function
(hence those factors are considered small compared to even just reading the input once).
For instance, consider the following case.
Let $f()$ denote some arbitrary computable function,
$Q$ the query,
$\inp$ the size of its largest input relation
and $\out$ the size of its output.
Then, a detailed complexity formula such as
$\O(f(|Q|)\cdot\inp^{f(|Q|)} + (\log\inp)^{f(|Q|)}\cdot\out)$ would simplify to
$\tilde\O(\inp^{f(|Q|)} + \out)$.
Note how the exponent that depends on $|Q|$ does not disappear in the first term
and how the entire poly-log factor disappears in the second. Whenever we want to analyze
performance differences at finer granularity, we will also show the detailed complexity formulas
in standard $\O$-notation.

All material will be self-contained, i.e., we only assume familiarity with fundamental database
concepts that would be covered in a typical undergraduate database course, and we do not
require previous knowledge of optimal join algorithms or top-\k queries.

\smallsection{Outline of the tutorial} 
This is a 90-minute tutorial consisting of three main parts:
\begin{enumerate}
    \item Top-\k algorithms for join queries
    \item (Worst-case) optimal join algorithms
    \item Ranked enumeration over join queries: optimality, ranking functions, and empirical
    comparison of the most promising approaches.
\end{enumerate}
We will conclude with a variety of open research problems.
Slides and videos of the tutorial will be made available on the tutorial web
page.\footnote{\url{https://northeastern-datalab.github.io/topk-join-tutorial/}}

\section{Part 1: Top-\k algorithms}

The first part of the tutorial presents core techniques for answering top-\k queries in databases
with a particular focus on those supporting joins~\cite{ilyas08survey,lin18topk},
while only briefly touching on top-\k
problems in other contexts such as
single-table queries~\cite{rahul19topk}
that often have a geometric nature~\cite{mouratidis17topk}.
In general, top-\k aims to prioritize input tuples that could contribute to any of the
\k top-ranked results over those that cannot, 
often pruning the latter as early as possible. 
Complications arise because the importance of a result tuple
(often captured by its aggregate weight)
typically depends on the weights of the input tuples that join to produce it.
This limits the choices of \emph{ranking functions} for which efficient computation and effective pruning
are possible.

One of the best-known top-\k approaches is the Threshold Algorithm (TA)~\cite{fagin03},
for which Fagin, Lotem and Naor received the 2014 G\"{o}del Prize, both for the algorithm's
simplicity and its strong instance-optimality guarantee.
Conceptually, TA operates on a single table that was partitioned vertically, with each
partition being managed by a different external service that knows the ranking only for its partition.
A middleware's challenge is then to combine those individual rankings to find the global winners
for the full table. TA's cost is measured in term of the number of tuple fragments retrieved by the
middleware from the external sources, but it \emph{does not take the actual join cost into account}.
This is acceptable in the target application, because TA supports only a very limited type of join,
also termed ``top-\k selection query''~\cite{ilyas08survey}, where tuple fragments from different
partitions join 1-to-1 on a unique object identifier to piece together a row of the full table.

TA marks the culmination of a series of papers where Fagin introduces the problem and
proposes an algorithm, now known as ``Fagin's algorithm'' (FA)~\cite{fagin96fa, fagin99fa, fagin98fa},
which does not have TA's strong optimality guarantees. 
FA also motivated several approaches that
essentially proposed TA before the famous TA paper, but without identifying and proving the algorithm's
instance optimality. This includes work by Nepal and Ramakrishna~\cite{nepal99ta} and
G{\"u}ntzer et al.~\cite{guntzer00qcombine}, 
the latter of which also incorporates heuristics for deciding which list to fetch
tuples from. 
TA in turn motivated various extensions of the idea to more general join problems, including
J*~\cite{natsev01},
Rank-Join~\cite{ilyas04},
LARA-J* \cite{mamoulis07lara},
a-FRPA \cite{finger09frpa},
SMART~\cite{wu10topk},
among others surveyed by Ilyas et al \cite{ilyas08survey}.
All these algorithms register significant performance gains when the \k top-ranked join results
depend on only ``a few'' top-ranked tuples from the input tables. In general, they attempt to minimize how
deep down the list they have to go in each pre-sorted input table 
until they can guarantee that the correct
\k results have been determined. To achieve the latter, they derive a bound on the score of possible join
results containing yet-unseen tuples and update this bound after accessing an input tuple.
Our intention here is to highlight the specific innovations introduced by each algorithm, which
mostly aim to navigate the tradeoff between cost for accessing tuples
\cite{schnaitter09depth,ilyas06rank}
and computing improved 
bounds~\cite{schnaitter08pbrj} for early termination.

Similar to TA, analytical results in this space are generally stated in terms of the number of input
tuples accessed. We revisit those results and analyze the algorithms in the standard RAM model of computation.
We are particularly interested in their worst-case behavior when some of the input tuples contributing to the
top-ranked result are at the bottom of an individual input relation and will generally explore to what degree
they suffer from large intermediate results, especially for cyclic joins.

\section{Part 2: Optimal Join Algorithms}\label{sec:optimal-join-algorithms}

The second part of the tutorial presents both classic and state-of-the-art results on optimal processing
of join queries, using minimal examples such as path, triangle, and 4-cycle queries in graphs.
We provide a brief summary of selected approaches that will be discussed.
In addition to algorithms, we will also take a closer look at various competing notions of
\emph{optimality}~\cite{khamis17panda,ngo2018worst}. For the following discussion,
recall that $\inp$ is the size of the largest input relation,
$\out$ is the size of the output 
and
we generally express complexity results in terms of data complexity
in $\tilde\O$-notation. Furthermore, the cost analysis
makes no assumption about the existence of pre-computed data structures on the input relations
at query-submission time, including any type of indexes or materialized views.
If the algorithm needs such a data structure, it has to create it
from scratch, i.e., this cost is reflected in the query time.

For a lower bound, notice that query $Q$ has to examine each input tuple at least once and has to write
out each result. This means that join evaluation has complexity at least
\[
\Omega(\inp + \out).
\]
Somewhat amazingly, the \emph{Yannakakis algorithm}~\cite{DBLP:conf/vldb/Yannakakis81} 
achieves
\[
\tilde\O(\inp + \out) 
\]
for \emph{acyclic} queries, essentially matching the lower bound.

Its secret of success is the property that after a \emph{full reducer} pass,
consisting of semi-join reductions~\cite{DBLP:journals/jacm/BernsteinC81}
between pairs of joining input relations,
the database is left in a state of global consistency \cite{DBLP:journals/ai/Dechter92}
, where any intermediate join result can be extended to a valid output tuple.

Unfortunately, as Ngo et al.~\cite{ngo2018worst} convincingly argue, for join queries with \emph{cycles} the
$\tilde\O(\inp + \out)$ bound is unattainable based on well-accepted complexity-theoretic assumptions. 
They therefore propose the notion of \emph{worst-case-optimal (WCO)}
join algorithms of time complexity
\[
\tilde\O(\inp + \out_{\textrm{WC}}),
\]
where $\out_{\textrm{WC}}$ denotes the size of the greatest possible output of query $Q$
\emph{over any database instance}.

For $\out_{\textrm{WC}}$, Atserias, Grohe, and Marx~\cite{AGM} provide a tight upper bound
by connecting join-output size to the \emph{fractional edge cover} of the corresponding query hypergraph.
This is now known as the AGM bound, and it is tight in the sense that there exist database instances for which the
output size indeed matches the bound. Follow-up work extended the AGM bound to general conjunctive queries
with projections and/or functional dependencies~\cite{gottlob12fds} as well as degree 
constraints~\cite{abo2016degree,khamis17panda}, which generalize the concept of functional dependencies.

A variety of WCO join algorithms have been proposed to match the AGM bound
\cite{Kalinsky16wco,
navarro19wco,
ngo2018worst,
Ngo:2014:SSB:2590989.2590991,
veldhuizen14leapfrog}.
In contrast to the common ``two-relations-at-a-time'' approach, i.e., binary join plans,
favored by database optimizers, they take a more
``holistic'' approach by computing a multiway join directly. Consider the often used
\emph{triangle query}, a natural join over input relations
$R(A, B) = S(B, C) = T(C, A) = \{(1,1), (2,1),\ldots, (n/2,1), (1,2), (1,3),\ldots, (1,n/2)\}$.
No matter the join order for a binary join plan,\footnote{The three possible join orders are:
$(R \bowtie S) \bowtie T$, 
$(R \bowtie T) \bowtie S$, or
$(S \bowtie T) \bowtie R$.} 
the first binary join produces $\O(n^2)$ intermediate results, even though the AGM bound
shows that final output size cannot exceed $n^{1.5}$.
As a consequence, the binary-join approach has complexity $\tilde\O(n^2)$, while a WCO join algorithm like
\GenJoin~\cite{Ngo:2014:SSB:2590989.2590991} 
or \NPRR~\cite{ngo2018worst}
computes the output in time $\tilde\O(n^{1.5})$.

Unfortunately, WCO join algorithms are not \emph{output-sensitive}, i.e., their complexity does not improve
for database instances resulting in small output. Consider again the triangle query.
If there are indeed $\Theta(n^{1.5})$ results, then $\tilde\O(n^{1.5})$ join time is the best one can
hope for. On the other hand, if there are zero triangles for a given database instance, then one would hope
to be able to achieve running time closer to $\tilde\O(n)$, i.e., the time it takes to read the input.
This applies also to the Boolean version of the query, which asks \emph{if} there are any triangles, but
does not need to return any of them. These issues are addressed by a different notion of optimality
that requires the join algorithm to have time complexity
\[
\tilde\O(\inp^d + \out)
\]
for the smallest value of parameter $d$ possible~\cite{khamis17panda}.
In contrast to WCO join algorithms, this complexity
depends on the output size on the given database instance, not the largest output over any database instance.
Here $d$ is a \emph{width} parameter that captures the ``degree of acyclicity'' of the join hypergraph.
Intuitively, the smallest possible $d$ for a given query establishes its \emph{intrinsic difficulty}.
For acyclic queries, $d=1$ and hence the ideal complexity $\tilde\O(\inp + \out)$ is achievable with the
Yannakis algorithm as we discussed above.

For cyclic queries, the situation is more complicated and different notions of width have been 
explored~\cite{GottlobGLS:2016}.
From a practical point of view, algorithms with $\tilde\O(\inp^d + \out)$ complexity all follow the same
high-level approach. They first decompose a cyclic join query into a tree-shaped acyclic join query
and materialize the derived relations needed as input for each tree node. Then they run the Yannakakis algorithm
on the acyclic join over the derived relations. The total time complexity is generally determined by the
size of the largest derived relation. We will survey the different decomposition methods that have been proposed 
\cite{GottlobLS:2002,
gottlob03width,
gottlob09generalized,
greco17greedy,
greco17consistency,
grohe14fhtw,
Marx:2013:THP:2555516.2535926,
ROBERTSON1986309}
and highlight their relationships. The current frontier has been established by the
\emph{submodular width}~\cite{Marx:2013:THP:2555516.2535926}.
Its key innovation, from a practical point of view,
is that it decomposes a cyclic query into a \emph{union of multiple trees}, each one receiving a subset of the input.
This enables lower widths compared to decompositions to a single tree. For example, on the 4-cycle query
both the WCO 
\GenJoin~\cite{Ngo:2014:SSB:2590989.2590991} 
and approaches based on single-tree decompositions have complexity
$\tilde\O(\inp^2)$, the former due to worst-case output size being quadratic in input size and the latter
due to the fractional hypertree width being $d=2$. In contrast, submodular width is 1.5 and hence
algorithms like PANDA~\cite{khamis17panda} that rely on decompositions into multiple trees
achieve complexity $\tilde\O(\inp^{1.5} + r)$, which is better for \emph{small output size} $\out = \O(\inp^{1.5})$.

Decomposition techniques for cyclic queries also play a role in \emph{factorised databases}, which aim to reduce
query complexity by cleverly representing (intermediate) results in a factorised format 
\cite{bakibayev12fdb,olteanu16record,olteanu12ftrees,olteanu15dtrees}. We will
survey the key insights of this line of work and then conclude this part of the tutorial with an overview of extensions
providing support for aggregates~\cite{abo16faq,bakibayev13fordering,khamis17juggling}. Due to time
constraints, we will only provide pointers to other exciting extensions, including those to
machine learning~\cite{khamis18ml,khamis18acdc,kumar15ml,olteanu16ml,schleich19ml,schleich16ml},
degree information~\cite{joglekar18degree,khamis17panda},
inequalities~\cite{AboKhamis:2019:FAQ:3294052.3319694},
negation~\cite{khamis19negation},
result compression~\cite{deep18compressed},
dynamic settings~\cite{idris19dynamic,idris20dynamic_theta,kara19triangles},
and approaches aiming for stronger notions of optimality~\cite{ngo14mine,Khamis:2016:JVG:3014437.2967101}.
It is also worth noting that some of these novel join algorithms have been implemented in prototype
systems for graph processing~\cite{aberger17emptyheaded,hogan19sparql,kalinsky20graph}.
A historical perspective on WCO join algorithms together with open problems in the area have recently
been summarized by Ngo~\cite{ngo18open}.

\section{Part 3: Ranked Enumeration over joins (``any-\k'')}

The third part of the tutorial focuses on \emph{optimal ranked enumeration} over both
acyclic and cyclic joins, which has started to attract attention 
recently~\cite{chang15enumeration,deep19,KimelfeldS2006,Tziavelis:fullversion,yang2018any,YangRLG18:anyKexploreDB}.
A ranked-enumeration algorithm returns the join results in the order of importance as imposed by a
ranking function. Its goal is to minimize the time for returning the $k$ top-ranked results
\emph{for every value of $k$}. Stated differently, the algorithm must return query results one-by-one
in ranking order without knowing the value of $k$ in advance. While some
top-\k approaches support this functionality or can easily be extended to do so,
others rely on knowing $k$ for pruning lower-ranked results. In order to more
clearly distinguish between them, we will refer to ranked-enumeration algorithms also as
``\emph{any-\k}'' join algorithms as a shorthand for ``\emph{anytime top-\k}.''

Despite being reminiscent of the general concept of an anytime algorithm~\cite{Zaimag96,Boddy:1991,DBLP:conf/sigmod/HeuvelIGGT19,DBLP:journals/vldb/FinkHO13},
any-\k algorithms are not approximating the query result~\cite{Mozafari:2017:AQE:3035918.3056098}.
Instead, they reside squarely at the intersection of top-\k and optimal joins, and we will
discuss how they are impacted by ideas from both. This tutorial will also highlight an interesting
connection to \emph{constant-delay join enumeration} 
algorithms~\cite{segoufin15constenum,Berkholz:2017:ACQ:3034786.3034789,bagan07constenum,DBLP:journals/tocl/DurandG07},
which produce all query results in quick succession after a short pre-processing phase,
albeit in no particular order. Specifically, if an algorithm returns join results with
constant delay after spending time $t_{\textrm{prep}}$ on pre-processing, then it guarantees join time
$\tilde\O(t_{\textrm{prep}} + \out)$ and hence gives an output-sensitive complexity guarantee.
It therefore would seem natural to extend such approaches to ranked enumeration by
investing ``a little more'' into the pre-processing phase in order to return the
results in the right order with constant or logarithmic 
delay in input size. (The latter is also $\tilde\O(1)$.)

The center piece of this part of the tutorial are recent results showing that any-\k algorithms,
for a very general definition of the join query, can be modeled as extensions of non-serial dynamic
programming (DP)~\cite{Tziavelis:fullversion}. This view reveals common foundations between a variety
of solutions for problems that had been studied in isolation, often re-inventing the wheel:
\k-shortest paths~\cite{eppstein16kbest} and their relationship to 
DP~\cite{bertsekas05dp,Cormen:2009dp,dpv08book},
graph-pattern search~\cite{chang15enumeration,yang2018any}, and earlier approaches to ranked enumeration
over joins~\cite{deep19,KimelfeldS2006}.
We will demonstrate how these approaches rely on two different major techniques to support the any-\k property.

The first is the \emph{Lawler-Murty procedure}~\cite{lawler72,murty1968} that has been used
in the database community to design algorithms for ranked enumeration~\cite{KimelfeldS2006}
and for graph-pattern search~\cite{chang15enumeration,yang2018any}.
After identifying the top-ranked result, it cleverly partitions the problem space in order to
find the second-best result as the best solution in one of those subspaces; then it recursively
proceeds by further partitioning that ``winning'' subspace.
A direct application of the procedure that solves each partition from scratch
leads to a delay that is polynomial in the size of the
input~\cite{KimelfeldS2006}. 
Similar attempts with polynomial-delay results 
have also been made for the equivalent Constraint Satisfaction Problem (CSP) \cite{DBLP:conf/cp/GrecoS11,DBLP:journals/jcss/GottlobGS18}.
However, it was recently shown that by exploiting the inherent structure of the join problem, the delay can be
reduced to $\O(\log k) = \tilde\O(1)$~\cite{Tziavelis:fullversion}.

The second approach for adding ranked-enumeration capabilities to a standard DP algorithm originates
from \k-shortest-path solutions
\cite{eppstein1998finding,
hoffman59shortest,
jimenez03shortest,
martins01kshortest}
and it relies on a recursive enumeration algorithm that exploits a
generalization of the DP principle of optimality~\cite{dreyfus69shortest,
bellman60kbest,
jimenez99shortest}.
The same recursive call structure appears to have been rediscovered
in recent work on ranked enumeration for conjunctive join queries~\cite{deep19}.

We will present recent theoretical and empirical evidence
\cite{Tziavelis:fullversion}
that neither of the two major 
approaches (Lawler-Murty vs.\ recursive enumeration) dominates the other.
In general, these deeper relationships between seemingly different problems and algorithms are fascinating
in their own right. Besides, we argue that they are essential for the design of optimal
ranked-enumeration algorithms over joins, including generalizations that go beyond natural
and Boolean conjunctive queries.

We conclude with an overview of interesting open problems at the intersection of
joins and top-\k queries.

\section{Author information}

\textbf{Nikolaos Tziavelis} is a PhD student at the Khoury College of Computer Sciences of Northeastern University.
His research interests lie in query processing and ranking problems.
He holds a MEng in Electrical and Computer Engineering from the National Technical University of Athens.

\textbf{Wolfgang Gatterbauer} is 
an Associate Professor at the Khoury College of Computer Sciences at Northeastern University.
His research interests lie in the intersection of theory and practice of data management
with a particular focus on uncertain and inconsistent data.
Prior to joining Northeastern, he was an Assistant Professor at Carnegie Mellon's Tepper School of Business,
and before that a PostDoc at University of Washington.
He received his PhD in Computer Science at Vienna University of Technology.

\textbf{Mirek Riedewald} is
an Associate Professor in the Khoury College of Computer Sciences at Northeastern University.
He received his PhD from the University of California at Santa Barbara and held positions as
Research Associate at Cornell University as well as visiting research positions at Microsoft Research
in Redmond and at the Max Planck Institute for Informatics (MPI-I) in Germany.
His research interests are in data management and analytics, with an emphasis on designing
scalable distributed analysis techniques for data-driven science. He has collaborated successfully
with scientists from many domains, including ornithology, physics, mechanical and
aerospace engineering, and astronomy.

\section{Acknowledgements}

This work was supported in part by %
the National Institutes of Health (NIH) under award number R01 NS091421 and by
the National Science Foundation (NSF) under award number CAREER IIS-1762268.
The content is solely the responsibility of the authors and does not necessarily represent
the official views of %
NIH or NSF.

\bibliographystyle{ACM-Reference-Format}
\balance
\bibliography{bibliography-anyk.bib}

\end{document}